\documentclass[aps,prb,twocolumn,groupedaddress,floats,showpacs]{revtex4}
\usepackage{latexsym}
\usepackage{dcolumn}
\usepackage[dvips]{graphicx}
\usepackage{amssymb}
\usepackage{graphics}
\usepackage{amsmath}
\usepackage{epsf}
\usepackage{bm}

\newcommand{\vecr}{{\bf r} }
\newcommand{\vecq}{{\bf q} }

\def\XXint#1#2#3{{\setbox0=\hbox{$#1{#2#3}{\int}$}
     \vcenter{\hbox{$#2#3$}}\kern-.5\wd0}}

\begin{document}
\author{Victor M. Galitski, Shaffique Adam and S. Das Sarma}
\title{Statistics of random voltage fluctuations and the
low-density residual conductivity of graphene}
\affiliation{Condensed Matter Theory Center, Department of Physics,
University of Maryland, College Park, MD 20742-4111}
\date{\today}
\begin{abstract}
We consider a graphene sheet in the vicinity of a substrate, which
contains charged impurities. A general analytic theory to describe
the statistical properties of voltage fluctuations due to the
long-range disorder is developed. In particular, we derive a
general expression for the probability distribution function of
voltage fluctuations, which is shown to be non-Gaussian. The
voltage fluctuations lead to the appearance of randomly
distributed density inhomogeneities in the graphene plane. We
argue that these disorder-induced density fluctuations produce a
finite conductivity even at a zero gate voltage in accordance with
recent experimental observations. We determine the width of the
minimal conductivity plateau and the typical size of the electron
and hole puddles.  We also propose a simple self-consistent
approach to estimate the residual density and the non-universal
minimal conductivity in the low-density regime. The existence of
inhomogeneous random puddles of electrons and holes should be a
generic feature of all graphene layers at low gate voltages due to
the invariable presence of charged impurities in the substrate.
\end{abstract}

\pacs{81.05.Uw, 72.10.-d, 73.40.-c}
\maketitle

\section{Introduction}
\label{Intro}

 Charge inhomogeneities are known to play an important role in
 understanding transport properties of semiconductors.
 The randomly positioned impurity ions give rise to a random
 electrostatic potential and to local modulations of the electron density.
The corresponding phenomena have been extensively studied in bulk
 three-dimensional semiconductors and two-dimensional semiconductor
 heterostructures.~\cite{kn:efros1993} Recently, it became clear that in
 graphene,~\cite{kn:novoselov2004}
 which is a zero-gap semiconductor, the effects of
 Coulomb disorder on transport are even more drastic.

A particularly interesting property of graphene transport, dubbed
the ``minimal conductivity,'' is a saturation of the conductivity,
which happens at a low gate voltage with a width of $\Delta V_g
\sim 0.5 ~eV$, which is strongly sample dependent. The minimal
conductivity value itself is controversial with both a universal
 conductivity~\cite{kn:novoselov2005} and a
sample-dependent non-universal minimal
conductivity~\cite{kn:zhang2005} being claimed in the experimental
literature.  Perhaps the most remarkable feature of this
low-density residual conductivity is the approximate saturation of
the minimum conductivity with a voltage width of $\Delta V_g$,
without it dropping to zero at $V_g=0$, as the density is lowered
from an approximately linear voltage dependence of the
conductivity at high density~\cite{kn:hwang2006c,kn:nomura2006a}.
Although there are transport theories of Dirac fermions in a
``white-noise'' disordered potential~\cite{kn:fradkin1986} or in
clean systems~\cite{kn:katsnelson2006}, which give a universal
conductivity at the Dirac point, both the measured value of the
(non-universal) minimal conductivity and the observed existence of
a saturation region around $V_g=0$ are in direct conflict with the
intrinsic Dirac point physics.

Recent experiments~\cite{Columbia} have convincingly proven that
the observed phenomenon of ``minimal conductivity'' occurs due to
charge disorder trapped in the SiO$_2$ substrate. Each charged ion
produces a potential in the graphene layer, which locally mimics a
gate voltage and thus leads to a non-zero density in the
corresponding region even if the applied gate voltage is zero
(i.e., at the ``Dirac point''). If the substrate is electrically
neutral, the charges in the substrate do produce local density
inhomogeneities of electrons and holes, but the total density is
zero. The physical picture is that of potential valleys that
define electron puddles and mountains that define conducting hole
puddles~\cite{kn:hwang2006c}.  For the case of charge neutrality,
the point of zero external gate voltage defines the percolation
threshold, in the sense that the areas where the hole and electron
densities are non-zero are exactly equal to one half of the total
area of the sample (which is the percolation threshold in two
dimensions by duality). By changing the external gate voltage, one
creates an excess of electrons or holes and only one type of
carrier percolates. As long as the external gate voltage is much
smaller than the typical voltage fluctuation due to charged
impurities, $V_{\rm g} \ll
 \sqrt{\overline{V^2}}$, the conductivity is mostly
 determined by the random voltage fluctuations and is almost
 independent of $V_{\rm g}$. In this case, the conductivity
dependence is expected to have a plateau
 of width $\sqrt{\overline{V^2}}$ and be symmetric with respect to
 $V_{\rm g} \to - V_{\rm g}$. The second scenario that should
 be considered is a substrate, which is not
electro-neutral (which is perhaps a more experimentally relevant
picture). In this case, the dependence of the conductivity on the
gate voltage is not symmetric with respect to $V_{\rm g} \to -
V_{\rm g}$, but is shifted left or right, depending on the
substrate's total charge. The percolation picture still holds, and
in two dimensions, the duality between the percolating cluster and
its shoreline guarantees that the critical gate voltage at which
the electrons stop percolating is the same as that at which the
holes begin to percolate. The transition happens at some point
$V_{\rm g} \sim \overline{V}$.  The width of the minimal
conductivity plateau is expected to be of order
$\sqrt{\overline{\delta V^2}}$.
 At gate voltages much larger than this value, transport is expected to be
 insensitive to the voltage fluctuations and is described in terms of
 the standard transport theory developed earlier by two of the
 authors~\cite{kn:hwang2006c}. We should mention that the
 picture outlined above is entirely {\em classical}.  At
 the lowest temperatures, the localization physics should become important.
 However, we believe that the current experiments are being done
 in the temperature regime where the quantum interference
 corrections are irrelevant - a
 quantitative agreement between the available experimental data at high
 carrier density
  and the Boltzmann theory of Ref.~[\onlinecite{kn:hwang2006c}] strongly
support this claim.

To describe the inhomogeneous density profile and transport near
the Dirac point in graphene, it is important to have a complete
description of the statistical properties of the corresponding
random electrostatic potential, which creates the inhomogeneous
structure in the first place. In this paper, we develop a
systematic analytic theory to derive the relevant probability
distribution functions of voltage fluctuations and correlators of
voltages due to the charge inhomogeneities. We note here that
similar potential fluctuations due to randomly distributed charges
have been considered previously in the literature in the context
of electron transport in semiconductor heterostructures. E.g.,
Refs.~[\onlinecite{kn:efros1993},\onlinecite{Nixon}] have studied
the statistics of potential and density fluctuations numerically
in the framework of a non-linear screening model. Below, we study
the voltage fluctuations analytically. Using the developed method,
we show that the PDF of voltage fluctuations is generally
non-Gaussian and derive explicit analytic expressions for the PDF
in various limits. Using the obtained analytical results, we
estimate the typical size of charge-disorder-induced electron/hole
puddles for a typical graphene sheet. The corresponding results
appear to be in a very good agreement with the available
experimental data.~\cite{GG} We also suggest a self-consistent
(mean-field-like) procedure to estimate the typical density at the
Dirac point and the remanent conductivity near the percolation
threshold. We also discuss the relation between the suggested
percolation-like picture of graphene transport near the Dirac
point and the usual diagrammatic transport theory, which works
well at high gate voltages.



\section{General method of calculating statistics of voltage fluctuations}
\label{General}

Let us consider a two-dimensional plane in the vicinity of a
substrate containing randomly positioned charged impurities.
Carriers in the 2D plane screen the bare Coulomb potential
produced by the charges in the substrate. The screening properties
and the corresponding effective potential depend on the nature and
the density of the carriers, but for the purpose of this section,
the exact form of the potential is not important. Let us denote it
as $v({\bf r})$.

To derive the PDF of voltage fluctuations, we exploit a method,
used e.g. by Larkin et al.~\cite{kn:larkin1971} in the context of
a mean-field approach to random spin systems. First, we define the
PDF as follows
\begin{equation}
\label{PDF_def} P[V] = \left\langle \delta \left[ V - \sum_i
v({\bm \rho}_i) \right] \right\rangle,
\end{equation}
where $V$ is the random voltage, the index $i$ in the sum labels
the charges, $v({\bm \rho}_i)$ is the potential produced by the
$i$-th impurity, and the angular brackets correspond to averaging
over all possible positions of impurities. Using the standard
representation of the $\delta$-function, we get
\begin{equation}
\label{PDF2} P[V] = \int \frac{d\lambda}{2 \pi} e^{i V \lambda}
\left\langle \prod_i \exp \left[ - i \lambda v({\bm \rho}) \right]
\right\rangle.
\end{equation}
At this point, we assume that the impurities are uncorrelated,
which allows us to exchange  the averaging operation with the
product over the impurities in Eq.~(\ref{PDF2}).  We obtain
\begin{equation}
\label{PDF3} P[V] = \int \frac{d\lambda}{2 \pi} e^{i V \lambda}
\left\langle \exp \left[ - i \lambda v({\bm \rho}) \right]
\right\rangle^{N_{\rm imp}},
\end{equation}
where $N_{\rm imp}$ is the number of impurities in the substrate.
Now, we average over all possible positions of impurities, which
have the equal probabilities. This implies simply evaluating an
integral over the entire two-dimensional area ${\cal A}$.
\begin{equation}
\label{<3>} \left\langle \exp \left[- i \lambda v({\bm \rho})
\right] \right\rangle = 1 -  \frac{1}{\cal A} \int\limits_{\cal A}
d^2\rho \left\{ 1 -  \exp \left[- i \lambda v({\bm \rho})
\right]\right\},
\end{equation}
where we performed a trivial operation of subtracting and adding a
unity. Using Eq.~(\ref{<3>}), we see that in thermodynamic limit,
the PDF  can be written as
\begin{eqnarray}
\label{PDF*} P[V] = \int  \frac{d\lambda}{2 \pi} e^{i V \lambda}
\exp \left[ n_{\rm imp} \int\limits d^2\rho \left\{ 1- \exp \left[
i\lambda v({\bm \rho}) \right] \right\} \right],
\end{eqnarray}
where $n_{\rm imp} = N_{\rm imp}/{\cal A}$ is the concentration of
impurities. Note that this result does not depend on the specific
form of the potential and is completely general. We emphasize here
that Eq.~(\ref{PDF*}) describes a random potential due to a
charged substrate, which contains disorder of just one type. If
the substrate contains charges of different types, at the level of
Eq.~(\ref{<3>}), we have to perform another averaging over the
distribution function of charges, $C[Q]$. In particular, if the
substrate is electroneutral and contains charges of two types of
opposite signs $\pm Q_0$, then the charge distribution function is
simply $C[Q] = \frac{1}{2} \left[ \delta(Q-Q_0) + \delta(Q+Q_0)
\right]$, which leads to the following PDF for the electrically
neutral substrate:
\begin{eqnarray}
\label{PDF0} P_0[V] = \int  \frac{d\lambda}{2 \pi} e^{i V \lambda}
\exp \left[ n_{\rm imp} \int\limits d^2\rho \left\{ 1- \cos \left[
\lambda v({\bm \rho}) \right] \right\} \right].
\end{eqnarray}

To describe the  distribution and sizes of disorder-induced
droplets, we also consider another PDF function, which determines
the distribution of voltages in two different points in the film.
Similarly to Eq.~(\ref{PDF_def}), we  define it as
\begin{equation}
\label{P[V1V2]} P[V_1,V_2,\rho_{12}] = \left\langle \delta \left[
V_1 - \sum_i v({\bm \rho}_{i1})\right] ~ \delta\left[ V_2 - \sum_j
v({\bm \rho}_{j2}) \right] \right\rangle,
\end{equation}
where ${\bm \rho}_{ij} = {\bm \rho}_i - {\bm \rho}_j$ is the
distance between the two points. This correlation function
(\ref{P[V1V2]}) determines the conditional probability of finding
a voltage $V_2$ in the point ${\bm \rho}_{12}$, provided that in
the origin, the voltage is equal to $V_1$. Thus, the decay of this
correlation function will determine the typical size of the charge
disorder-induced puddles of electrons or holes.

Following the simple steps, which led to Eq.~(\ref{PDF*}), we
obtain for Eq.~(\ref{P[V1V2]})
\begin{eqnarray}
\label{PDF12*} && P[V_1,V_2,\rho_{12}] =
\int \frac{d\lambda_1 d\lambda_2}{(2 \pi)^2} e^{i V_1 \lambda_1 - i \lambda_2 V_2}\\
&& \times  \exp \left[ n_{\rm imp} \int\limits d^2\rho_0 \left\{
1- \exp \left[i \lambda_2 v({\bm \rho}_{20}) - i \lambda_1 v({\bm
\rho}_{01}) \right] \right\} \right]. \nonumber
\end{eqnarray}

Formally, Eqs.~(\ref{PDF*}) and (\ref{PDF12*}) completely describe
the probability distribution of voltages for a given effective
interaction, $v({\bf r})$. However, exact analytical calculation
of the full PDFs is possible only in the simplest cases of pure
Coulomb interaction and large-distance asymptote of the linearly
screened Coulomb (see below Sec.~\ref{Models123}). To correctly
interpolate between the two behaviors, one has to consider the
general form of the screened interaction potential.  The latter
has a relatively complicated structure in real space and generally
it is not possible to calculate the full PDF analytically.
However, it is possible to evaluate exactly the relevant moments
of the PDF. These moments can be calculated from Eqs.~(\ref{PDF*})
and (\ref{PDF12*}), by simply evaluating derivatives with respect
to the auxiliary parameter $\lambda$,
\begin{eqnarray}
\overline{V^n} =\!\!\!\int \frac{d\lambda}{2 \pi} e^{ - n_{\rm
imp} \int d^2\rho \left[ 1 - e^{-i\lambda v(\rho)} \right]} \left(
- i \frac{\partial} {\partial \lambda} \right)^n \!\!\!\int dV
e^{i\lambda V}. \label{<V>}
\end{eqnarray}
In a similar way, one can determine the real space
correlation function of the random voltages
\begin{eqnarray}
 \overline{\left[V(\vecr) - \overline{V}\right]
 \left[V(0) - \overline{V}\right]}= n_{\rm imp}
\int \frac{d^2q}{\left( 2 \pi \right)^2}\tilde{v}^2(q) e^{i \vecq
\cdot \vecr}, \label{Eq:VV}
\end{eqnarray}
which is valid for any effective potential $\tilde{v}(q)$.  

\section{Derivation of PDF for specific model interactions}
\label{Models123}

In this section,we demonstrate the application of the method of
Sec.~\ref{General} by considering three specific model forms of
the interaction $v(\rho)$.

\subsection{PDF of voltage fluctuations for bare Coulomb
potential} \label{model1}

 We start with the bare Coulomb interaction $v_0(\rho)
= Q_0/\rho$, where $\pm Q_0$ are the charges of the impurities. We
have to evaluate the following integral [see Eq.~(\ref{PDF*})]:
$$
I_0(\lambda) = 2 \pi n_{\rm imp} \int\limits_{0}^{\infty} d \rho
\rho \left[ 1- \cos \left(\frac{\lambda Q_0}{\rho}\right)
\right].
$$
Apparently, this integral diverges at large distances. So, we
expand the cosine term and get $\label{I2} I_0(\lambda) = \pi
n_{\rm imp} \lambda^2 Q^2 \int\limits_{\rho_{\rm min}}^{\rho_{\rm
max}} {d\rho}/{\rho} = \pi n_{\rm imp} \lambda^2 Q_0^2
\ln{\left({\rho_{\rm max}}/{\rho_{\rm min}}\right)}.$ To
regularize the divergences in the integral, we note that the
minimal distance can not be smaller than the distance from the
substrate. On the other hand, the large-distance divergence is
connected with the long-range nature of the Coulomb interaction
and is regularized by screening characterized by a Thomas-Fermi
screening wave-vector $q_{\rm s}$.  Summarizing, we write the PDF
corresponding to the pure Coulomb interaction
\begin{eqnarray}
\label{PDF_pure1} P_{\rm bare}[V] = \int  \frac{d\lambda}{2 \pi}
\exp\left[i V \lambda - \alpha(\lambda) \lambda^2 \right],
\end{eqnarray}
with
$$ \alpha(\lambda) \sim \pi n_{\rm imp} Q_0^2 \ln\left[ q_{\rm
s}^{-1} {\rm max}~\left\{ d,~|\lambda| Q_0\right\}\right].
$$
Note that this coefficient depends on $\lambda$ very weakly. We
can substitute $\lambda$ with a typical $\overline{\lambda} \sim
\max~\left\{\overline{V}^{-1},\alpha^{-1/2}\right\}$. After this
simplification the integral (\ref{PDF_pure1}) can be calculated
and we find a {\em Gaussian PDF} of voltage fluctuations with the
variance $\overline{V^2} = 2 \alpha$. We see that the typical
voltage is enhanced as compared with the na{\"\i}ve estimate,
$n_{\rm imp}^{1/2}Q_0$, by a large logarithmic factor.

\subsection{PDF of voltage fluctuations for large-distance
asymptote of screened interaction in 2D} \label{model2}

Another model interaction, which allows analytic treatment is the
large-distance asymptote of the screened interaction. Due to a wek
screening in two dimensions, it decays as a power law
$$
v_{\rm scr}({\bm \rho}) = {4 Q_0 }/ {q_{\rm
s}^{2}\rho^3}.
$$
To determine the statistics of the random voltage, we need to
evaluate the following integral [see Eq.~(\ref{PDF*})]
$$
I_{\rm scr}(\lambda) = 2 \pi n_{\rm imp} \int\limits_{0}^{\infty}
d \rho \rho \left[ 1- \cos \left(\frac{4 \lambda Q_0}{q_{\rm
s}^{2} \rho^3}\right) \right].
$$
This integral does not have any infrared problems due to
screening. After some algebra, we find the following PDF
\begin{equation}
\label{PDF_scr} P_{\rm scr} [V] = \frac{1}{\pi \beta}
\int\limits_0^\infty dx \cos\left(\frac{V}{\beta} x\right)
e^{-x^{2/3}}, \end{equation} where $\beta = n_{\rm imp}
[\Gamma(1/3) \pi/2]  \left(4 Q_0 /q_{\rm s}^2 \right)^{2/3}$ is a
parameter, which depends on the carrier density and $\Gamma(z)$ is
the Gamma function. From Eq.~(\ref{PDF_scr}), one can see that in
the case of a screened interaction {\em the PDF of voltage
fluctuations is strongly non-Gaussian}.  One can check that all
even moments of this PDF diverge (all odd moments are trivially
zero due to assumed electroneutrality), which corresponds to a
singular behavior of the model potential at small distances.
Obviously, this divergence is an artifact of our approximation
about the interaction potential, which is justified only at large
distances.

\section{Screened Coulomb potential}
\label{model3}

To correctly interpolate between the small distance (large-$V$)
asymptotics of Sec.~\ref{model1} and large-distance (small-$V$)
behavior of Sec.~\ref{model2}, we need to consider a general form
of the effective interaction, which includes both screening and a
finite width of the spacer layer.  Below, we specialize to the
case of linear screening, where we use the following form of the
effective potential
\begin{equation}
\label{vscr} \tilde{v}(q) = \frac{2 \pi Q_i e^{-qd}}{q + q_{\rm
s}},
\end{equation}
where $q_{\rm s}^{-1}$ is the Thomas-Fermi screening length and
$d$ is the distance between the 2D graphene plane and the
substrate. Note that the screening length is graphene is
density-dependent. If there is no external gate voltage, there is
no density to start with. However, it inevitably appears due to
the random voltage fluctuations and can be determined
self-consistently in the framework of a mean-field-like theory
(see Sec.~\ref{General} and Ref.~[\onlinecite{AHGD}]). In this
section, we simply take the quantity $q_s$ as a free parameter of
the theory.

\begin{figure}
\centering
\includegraphics[width=3.5in]{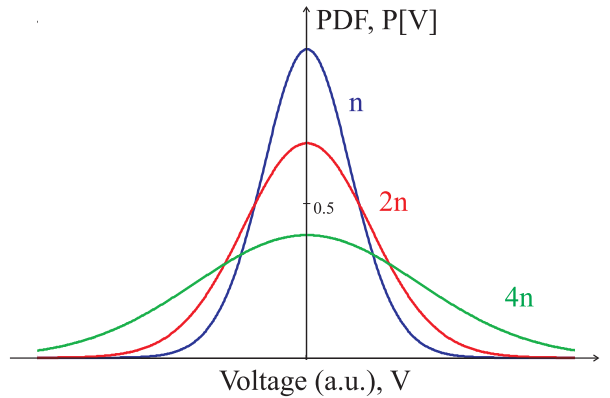}
\caption{\label{Fig:PDF}~(Color online) This figure shows the
representative probability distribution functions of random
voltages arising from the screened potential~(\ref{vscr}) of
randomly distributed charges in an electrically neutral substrate
for three different densities (other parameters are fixed).}
\end{figure}

Due to a complicated structure of the potential in real space, it
is not feasible to calculate the full PDF (\ref{PDF*})
analytically. However, it is easy to perform a numerical
evaluation of the corresponding integrals. A result of such
calculation is shown in Fig.~\ref{Fig:PDF}, where representative
PDFs are plotted for three different densities. We note that
different portions of the curves can be very well-fitted with
Gaussians, but the entire dependence is definitely more
complicated than a simple Gaussian.  Moreover, if $\overline{V}
\ne 0$ (a charged substrate), the distribution function is not
even symmetric with respect to $\delta V = \left( V - \overline{V}
\right)$, i.e. $P[\delta V] \neq P[-\delta V]$.

Even though the full PDF can not be determined analytically, one
can evaluate explicitly its moments using Eqs.~(\ref{<V>}) and
(\ref{Eq:VV}). In particular, we calculate the second moment for
the case of Thomas-Fermi screening, $\overline{\delta V^2} =
\overline{\left( V - \overline{V} \right)^2}$; the result is
\begin{eqnarray}
\label{<VV>2} \overline{\delta V^2} = 2 \pi n_{\rm imp} Q_0^2
\left[ {\rm E_1}\, (2q_{\rm s}d) e^{2q_{\rm s} d} \left( 1 +2
q_{\rm s} d \right) - 1 \right],
\end{eqnarray}
where $E_1(z) = \int_z^\infty t^{-1} e^{-t} dt$ is the exponential
integral function. Obviously, the statistical properties of the
potential strongly depend on the parameter $q_{\rm s} d$.  Below
we give asymptotic expressions for $\overline{\delta V^2}$ in the
limits of large and small $q_{\rm s} d$
\begin{equation}
\label{P0} \overline{\delta V^2} = 2 \pi n_{\rm imp} Q_0^2 \left\{
\begin{array}{ll}
\left(2 q_{\rm s} d\right)^{-2},  & \mbox{if }\, q_{\rm s} d \gg 1;\\
\ln{\left[ \frac{1}{2 e \gamma q_{\rm s} d} \right]}, & \mbox{if
}\, q_{\rm s} d \ll 1,
\end{array}
\right.
\end{equation}
where $\gamma \approx 1.781$ is the Euler's constant, $e \approx
2.718$ is the base of the natural logarithm. We note that the last
expression essentially reproduces the result obtained above for
the bare Coulomb potential [see Eq.~(\ref{PDF_pure1}) and the text
below it].

\begin{figure}
\centering
\includegraphics[width=2.5in]{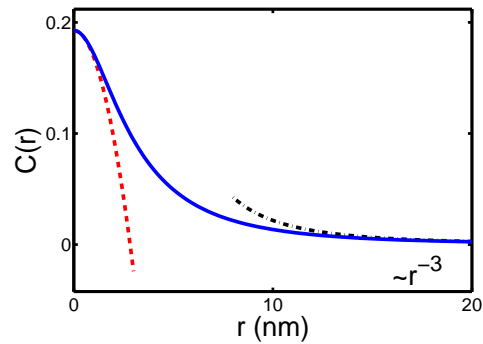}
\caption{\label{Fig:CorFun}~(Color online) The correlation
function $C(r)$ [see Eq.~(\protect\ref{Eq:VV}) and text for the
definition] for the charge impurity model with linear Thomas-Fermi
screening. The solid line shows numerical integration for $q_{\rm
s} d = 0.5$, while the dashed lines correspond to the asymptotic
analytical expressions for small and large $q_{\rm s} r$. The
chosen value of $q_{\rm s} d = 0.5$ is determined by the density
of carriers, where the experimentally observed behavior deviates
from the predictions of the classical Drude-Boltzmann theory.}
\end{figure}

Next we study the correlation function $\langle V({\bf r}) V(0)
\rangle$, which is related to the typical size of electron/hole
puddles. To simplify the notation we define a new function $C(r)$
according to the following relation $\langle V(\vecr) V(0) \rangle
= \overline{V}^2 + 2 \pi n_{\rm imp} Q_0^2 C(r)$. Using
Eq.~(\ref{Eq:VV}), one can calculate analytically all coefficients
in the Taylor series for the correlator $C(r)$
\begin{equation}
C(r) = C_0(2 q_{\rm s} d) + \sum_{m>0} (-1)^m (q_{\rm s} r)^{2m}
\frac{(2m-1)!!}{2^m m!} C_m (2 q_{\rm s} d),
\end{equation}
where $C_m(x) = [1/(2m)!](\partial_x^{2m+1})[x e^x E_1(x)]$. Note
that  $C_0$ reproduces the expression in the square brackets in
Eq.~(\ref{<VV>2}). In the limit  $q_{\rm s} r \ll 1 $, the
correlator behaves as $C(r) \approx C_0 - (C_1/2)  (q_{\rm s}
r)^{2}$. In the opposite limit of $q_{\rm s} r \gg 1$, one finds
that the correlation function decays as a power law $C(r) \sim 2
|1 - q_{\rm s} d|(q_{\rm s} r)^{-3}$. The general form of the
correlator has been calculated numerically in
Sec.~\ref{size.matters} and is shown in Fig~\ref{Fig:CorFun}.

\section{Application to graphene}
\label{observe}

\subsection{Size of the puddles}
\label{size.matters}

 Now, we  apply the general results obtained above
specifically to the case of a graphene sheet in the experimentally
relevant parameter regime. We estimate the density of charged
impurities, the parameter $q_{\rm s} d$, which determines the
behavior of the correlation function $C(r)$, the typical size of a
fluctuation-induced electron/hole droplet due to the charge
impurity fluctuations, and the typical width of the region where
the conductivity is expected to be almost a constant. We estimate
these quantities for typical graphene samples with mobilities $\mu
\sim 10^4 \mathrm{cm}^2/Vs$. The deviation from the semi-classical
Drude-Boltzmann theory occurs at carrier densities $n \sim
n_\mathrm{imp} \approx 5 \times 10^{11} \mathrm{cm}^{-2}$
~\cite{kn:hwang2006c}.  
This leads to the density of states $\nu_0 \approx  10^{10}
\mathrm{cm}^{-2} \mathrm{meV}^{-1}$ and the Fermi energy $E_{\rm
F} \approx 80 ~\mathrm{meV}$.  Assuming that the distance between
the graphene plane and the substrate is $d \sim 1 \mathrm{nm}$, we
find $q_{\rm s} d \sim 0.5$. The corresponding correlation
function is plotted in Fig.~2. This correlation function
determines the size of a disorder-induced puddle, which can be
roughly estimated as $l \sim 10 \mathrm{nm}$. We also can
determine the average value of the random potential as
$\overline{V} = n_\mathrm{imp}/\nu_0 \approx 0.05 ~\mathrm{eV}$
and $\sqrt{\overline{\delta V^2}} \approx 0.1 ~\mathrm{eV}$. The
former number determines the asymmetry in the conductivity
dependence, $\sigma(V_{\rm g})$, while the latter estimates the
width $\overline{\Delta V_g}$ of the constant conductivity
plateau. These numbers are in a reasonable agreement with the
available experimental data and validate the basic model that the
minimum conductivity in graphene is connected with the density
inhomogeneities caused by the charged impurities in the
$\mathrm{SiO}_2$ substrate.  We note two direct experimental
consequences of our theoretical findings: (1)~Dirtier samples
should have larger vales of $\Delta V_g$; implying  higher (lower)
mobility samples should have smaller (larger) values of $\Delta
V_g$.  (2)~In addition, $\Delta V_g$ has a nontrivial dependence
on $d$, the impurity location [see, Eq.~(\ref{P0})], which can, in
principle, be experimentally verified.

While the voltage (or density) width of the residual minimal
conductivity regime is explained naturally in our theory of
impurity induced inhomogeneous electron-hole puddle picture,
getting the exact values of the residual density and the minimal
conductivity will require a detailed  calculation involving
percolation through a network of random 2D electron and hole
puddles.~\cite{AHGD,CFAA} The following section  provides
estimates of these quantities in the framework of a
self-consistent ``mean-field'' approach, which should be
quantitatively reliable away from the precise percolation point.

\subsection{Mean-field approach and relation to diagrammatics}

In this section, we briefly outline possible approaches to relate
the voltage fluctuations to observables and, in particular,
discuss the regime of high gate voltages, where the usual
diagrammatic perturbation theory becomes quantitatively applicable
and show a possible relation between the discussed charge disorder
fluctuation phenomena and the diagrammatic approach. We also
formulate here a self-consistent mean-field-like approximation,
which allows us to roughly estimate some observable parameters in
the regime where perturbation theory breaks down.

While it is intuitively clear that the local voltage fluctuations
 lead to a non-vanishing density in the two-dimensional plane,
 the exact relation between the voltage distribution and the
density is  non-trivial. Generally, one  has to solve the
many-body Schr{\"o}dinger equation taking into account the
long-range disorder potential and Coulomb interaction between
carriers, which is an extremely complicated if not impossible task
to accomplish. Efros et al.~\cite{kn:efros1993} have previously
formulated a phenomenological approach in which the density is
determined from minimization of the energy of the system written
in terms of the position-dependent density $\Omega=\int d^2r
\left\{ \frac{1}{2} V({\bf r}) \left[ n({\bf r}) - \overline{n}
\right] + E[n({\bf r})] \right\}$, where the first term describes
the Coulomb interaction between the impurity charges and the
induced carriers and the local density, while the second term
represents the energy of a uniform liquid (which includes the
kinetic and exchange correlation effects). Formally, this approach
would allow to interpolate between the regimes of large and small
carrier density; but in the former case, i.e., if the chemical
potential is large, the effect of charge disorder fluctuations is
a small perturbation and the aforementioned phenomenological
approach should reduce to the usual microscopic perturbation
theory. E.g., one can determine the relevant density-density
correlation function $\left\langle \delta n({\bf r}) \delta n({\bf
r}') \right\rangle)$ by considering diagrams in Fig.~3a, where the
dashed line represents the electron scattering off of the Coulomb
impurity potential  and the shaded vertices imply a ladder of such
charge impurity lines. We note that to avoid infrared
divergencies, one has to consider screening of the impurity
potential by carriers, $U({\bf r},{\bf r}') = v({\bf r} -{\bf r}')
+ [v * \Pi * U] ({\bf r},{\bf r}')$ (where $v$ is the bare
potential and $\Pi$ is a polarization operator). Strictly
speaking, the screening properties and the interaction between two
charges are random quantities themselves, which depend on the
positions of the charges. In particular, the polarization
operator, which determines the screening properties is a random
matrix, $\Pi({\bf r},{\bf r}')$, which depends on two coordinates.
The probability distribution function of this matrix (in the
Gaussian approximation) is a functional $F[\delta\Pi] = \exp
\left[ -(1/2) \int_{1,2,3,4} \delta\Pi(1,2) C^{-1}(1,2;3,4)
\delta\Pi(1,2) \right]$, where $C^{-1}(1,2;3,4)$ is the inverse
correlator $\langle\delta\Pi({\bf r}_1,{\bf r}_2) \delta\Pi({\bf
r}_3,{\bf r}_4)\rangle$, which is described by the diagrams in
Fig.~3b. Clearly in the high-density regime, the average screening
length $q_s^{-1}$ will be determined just by the average density,
with small ``mesoscopic'' corrections. Similarly in this regime,
where the Fermi wave-length is the smallest length scale in the
problem, the local density in a particular point will be
determined just by the local voltage which can be viewed as a
local chemical potential in this point. Therefore, the average
density fluctuation (up to a numerical constant of order one)
should be of the order of $\sqrt{\langle (n - \overline{n})^2
\rangle} \sim n(\mu \sim \sqrt{\overline{\delta V^2}})$ [where
$n(\mu)$ is the dependence of the density on the chemical
potential in the uniform case and $\overline{\delta V^2}$ is given
by (\ref{<VV>2})].

\begin{figure}
\centering
\includegraphics[width=2.8in]{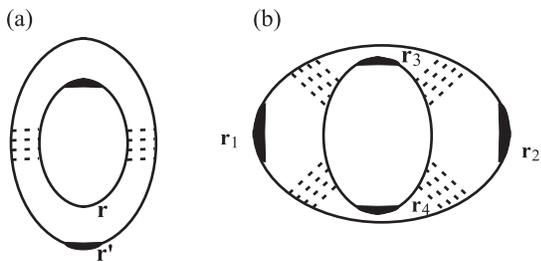}
\caption{\label{Fig:UCF}~Fig.~3a is the pictorial representation
of the disorder-induced density fluctuations $\left\langle\delta
n({\bf r}) \delta n({\bf r}')\right\rangle$, where the dashed line
represents the Coulomb impurity potential and the shaded vertex is
a ladder of these impurity lines. Fig.~3b is a diagram, which
contributes to the correlator $\langle\delta\Pi({\bf r}_1,{\bf
r}_2) \delta\Pi({\bf r}_3,{\bf r}_4)\rangle$ of the polarization
operators, which determine the screening properties and the
effective interaction (see text).}
\end{figure}

When the external gate voltage and thus the chemical potential and
the Fermi-momentum decrease, the microscopic perturbation theory
of density fluctuations in graphene breaks down. The Green's
functions and the screening properties of the impurity potential
in Fig.~3 both depend on the average density, which vanishes  at
the Dirac point and the corresponding computation becomes
ill-defined. However, it is possible to formulate a reasonable and
simple self-consistent scheme, which would allow to get an
order-of-magnitude estimate of the density in the vicinity of the
Dirac point. Namely, one assumes that the Green's functions and
the screening length correspond to a uniform density
$\sim\sqrt{\overline{\delta^2n}}$. Then, using these objects one
can calculate the density fluctuation, which, as argued above,
should be of order
$$
\sqrt{\overline{\delta^2n}} \sim n(\mu \sim \sqrt{\overline{\delta
V^2}}).
$$
But in this equation, the right hand-side of this
relation itself depends on $\overline{\delta^2n}$ via the
screening length $q_s^{-1}$ [see, Eq.~(\ref{<VV>2}) of
Sec.~\ref{model3}]. This leads to an algebraic non-linear
self-consistency equation for the typical density fluctuation. The
corresponding calculation has been done by us and is described in
detail in Ref.~[\onlinecite{AHGD}] (the corresponding estimate for
the residual density is $n \sim 0.1 - 1 n_{\rm imp}$).

\begin{figure}
\centering
\includegraphics[width=3.4in]{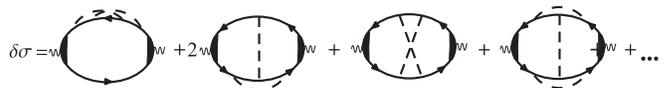}
\caption{\label{Fig:cond}~The first few diagrams in the
perturbative expansion of the conductivity. The dashed lines
correspond to the scattering off of Coulomb disorder potential.}
\end{figure}

A more important experimentally measurable quantity is the
conductivity. At very high gate voltages, the usual semiclassical
Drude-Boltzmann theory works perfectly well. As the chemical
potential is lowered, the ``mesoscopic fluctuation'' effects start
to appear. In the regime, when these effects are a small
correction to the classical conductivity, one should be able to
obtain them in the framework of the conventional diagrammatic
technique. This implies going beyond the Drude approximation (no
crossed impurity lines) and calculating ``quantum corrections'' to
the conductivity. A formal calculation would select the maximally
crossed weak-localization diagram, which  diverges in the
zero-temperature limit. However, the flattening of the
linear-in-$V$ conductivity most likely has nothing to do with the
weak localization or any other quantum interference effect of this
type. The effect of minimum conductivity is ``quantum'' only in
the sense that diagrammatically it appears beyond the classical
Drude approximation and is due to diagrams with some crossed
impurity lines. We note that in the case of long-range Coulomb
disorder, there are other dimensionless parameters apart from $k_F
l$ (namely, $r_s$ which formally should be assumed small for
perturbation theory to hold). If the dephasing time is short
enough, other types of diagrams (e.g., impurity vertex
corrections) may be more important than the weak localization
diagram. We believe that in the recent experiments, which observed
the minimum conductivity, the latter scenario is realized. The
fate of the weak localization correction is a very interesting but
separate question, which seems to be irrelevant to the effect of
the flattening of conductivity at low gate voltages. The initial
deviation of the $\sigma(V_g)$-dependence from the straight line
can be obtained from the usual perturbation theory by considering
just the next-to-leading diagrams with a finite number of crossed
Coulomb impurity lines (see Figs.~4b and 4c). The corresponding
calculation will be published elsewhere.

As noted above, in the limit of zero or very small gate voltage
(in the very vicinity of  the Dirac point), the microscopic
perturbation theory breaks down and only qualitative descriptions
are available. In this regime, transport occurs via a percolation
network of electron and hole puddles. Cheianov et al.~\cite{CFAA}
recently proposed a random resistor network model and calculated a
scaling behavior of the conductivity near the percolation
threshold. While the corresponding scaling function may be
universal, the value of the minimal conductivity itself is
non-universal (e.g., it depends on the choice of the underlying
lattice) and can not be determined within a random resistance
network model. The possibility that disorder associated with
random charged impurities could lead to a non-universal graphene
minimum conductivity was pointed out in
Ref.~[\onlinecite{kn:hwang2006c}], and the value of the minimal
conductivity was estimated in Ref.~[\onlinecite{AHGD}], by
assuming that it is of the order of Drude-Boltzmann conductivity
with the density determined by the value of the typical density
obtained via  the self-consistent procedure explained above. We
note that this estimate agrees very well with the available
experimental data~\cite{Columbia} (e.g., both the high-voltage
Drude behavior and the minimal conductivity can be determined
using just one fitting parameter - the density of impurities in
the substrate).

\section{Conclusion}
\label{Conclusion}

In this paper, we developed a simple and general method of
calculating statistical properties of voltage fluctuations due to
charge disorder in the vicinity of a conducting two-dimensional
plane. We used these results to describe the low-density
properties of a graphene sheet near a substrate with Coulomb
impurities. In particular, we estimated the typical size of the
electron and hole puddles and the width of the minimum
conductivity plateau. We also proposed a self-consistent
approximation scheme to estimate the typical residual density and
conductivity at the Dirac point. The corresponding results appear
to be in a good agreement with experimental data.~\cite{Columbia}

While our theory based on density fluctuations by extrinsic
charged impurities invariably present in the substrate provides a
reasonable mean-field description  of the observed graphene
non-universal conductivity minimum plateau around the charge
neutrality point, an important open question remains on the
 transport behavior precisely at the percolation
critical point $V_g = V_c$ where infinite electron and hole
percolation paths open for transport throughout the system.
Neither the mean-field theory of Ref.~[\onlinecite{AHGD}] nor the
perturbative diagrammatic approach can qualitatively access the
critical point where  the density fluctuations are very large.
Unfortunately, the
 problem of determining the minimum conductivity at $V_c$ remains
 inaccessible to the theory. We should mention that recently Cheianov et
 al.~\cite{CFAA} studied the critical percolation problem  by
 effectively including tunneling between the neighboring electron/hole
 puddles within a lattice random-resistance model, however their
 numerical analysis has not led to any estimate of the non-universal
non-universal conductivity at the percolation critical point.
Experiment seems to indicate a smooth and continuous behavior as
one goes from the $V_g < V_c$ to the $V_g > V_c$ regime with the
minimum conductivity remaining approximately a constant over a
finite interval of voltage $\delta V_g$, i.e. in the regime $V_g -
\delta V_g < V_c < V_g + \delta V_g$, as the voltage swept through
the charge neutrality point. The ``mean-field theory'' seems to
provide a reasonable quantitative estimate of both $\sigma_{\rm
min} \sim \sigma (V_c)$ and the value of $\delta V_g$ itself as
well as the conductivity $\sigma(|V_g|) \gg |V_c|$ far from the
electron-hole puddle percolation regime. Whether there is some
remnant universal quantum conductivity behavior, arising from the
special quantum-mechanical properties of the chiral Dirac-like
graphene band dispersion, remains an important open question. The
smooth and continuous experimental behavior of graphene transport
through the charge neutrality point empirically argues against any
universal behavior, but we can not rule it out on the current
theoretical grounds. In fact, closely connected with the critical
percolation transport is the question of (the absence of any
observed) Anderson localization in graphene. The strong disorder
associated with the random charged impurities should lead to
strong Anderson localization in graphene. Experimentally, however,
no such strong localization effect has ever been seen in graphene
and even the most disordered graphene samples seem to have a
finite minimum conductivity which is relatively temperature
independent around $T \lesssim 100$~mK. Why this is so is not
understood theoretically. It is possible that the crossover to
strong localization occurs at very low temperatures but more
experimental and theoretical work is needed to settle this
important question. The issues of the critical percolation
transport near the charge neutrality point and the apparent
absence of Anderson localization in graphene are both beyond the
scope of this work.

An important conclusion, which follows from our work is that there
will be no percolation induced metal-insulator transition in
graphene as there is in 2D semiconductor
systems~\cite{kn:dassarma2005b} and the graphene layer will always
conduct at low density (leading to a residual conductivity), as
long as there is no Anderson localization, simply because the
electron and hole percolation densities are exactly equal by
duality in two dimensions, and therefore, either the electron or
the hole channel always conducts independent of how strong the
impurity-induced inhomogeneous voltage fluctuations are. This
residual conductivity is not universal and depends on the density
of charge impurities.

Finally, we emphasize that the theoretical technique we develop in
this paper for calculating the non-Gaussian probability
distribution function of charged impurity induced voltage
fluctuations is completely general, and can be used for other 2D
systems, e.g. modulation-doped semiconductor heterostructures and
quantum wells, where such fluctuation-induced density
inhomogeneity and puddle formation effects \cite{kn:dassarma2005b}
are known to be important, often leading to percolation-induced
conductor-to-insulator transition. We believe that the theoretical
framework developed in this work can be adapted to 2D disordered
semiconductors providing insight into the nature of transport in
these systems.

Note added: After the first version of this manuscript had been
submitted, we received a preprint~\cite{kn:martin2007}, which
reported an explicit experimental observation of the electron-hole
puddles in graphene and provided an estimate of a puddle size in
quantitative agreement with our theory. Acknowledgements: This
work is supported by US-ONR.


\vspace*{-0.5cm}

\begin{thebibliography}{11}
\expandafter\ifx\csname natexlab\endcsname\relax\def\natexlab#1{#1}\fi
\expandafter\ifx\csname bibnamefont\endcsname\relax
  \def\bibnamefont#1{#1}\fi
\expandafter\ifx\csname bibfnamefont\endcsname\relax
  \def\bibfnamefont#1{#1}\fi
\expandafter\ifx\csname citenamefont\endcsname\relax
  \def\citenamefont#1{#1}\fi
\expandafter\ifx\csname url\endcsname\relax
  \def\url#1{\texttt{#1}}\fi
\expandafter\ifx\csname urlprefix\endcsname\relax\def\urlprefix{URL }\fi
\providecommand{\bibinfo}[2]{#2}
\providecommand{\eprint}[2][]{\url{#2}}

\bibitem[{\citenamefont{Efros et~al.}(1993)\citenamefont{Efros, Pikus, and
  Burnett}}]{kn:efros1993}
\bibinfo{author}{\bibfnamefont{A.~L.} \bibnamefont{Efros}},
  \bibinfo{author}{\bibfnamefont{F.~G.} \bibnamefont{Pikus}}, \bibnamefont{and}
  \bibinfo{author}{\bibfnamefont{V.~G.} \bibnamefont{Burnett}},
  \bibinfo{journal}{Phys. Rev. B} \textbf{\bibinfo{volume}{47}},
  \bibinfo{pages}{2233} (\bibinfo{year}{1993});
\bibinfo{author}{\bibfnamefont{B.~I.} \bibnamefont{Shklovskii}}
  \bibnamefont{and} \bibinfo{author}{\bibfnamefont{A.~L.} \bibnamefont{Efros}},
  \emph{\bibinfo{title}{Electronic Properties of Doped Semiconductors}}
  (\bibinfo{publisher}{Springer, New York}, \bibinfo{year}{1984}).


\bibitem[{\citenamefont{Novoselov et~al.}(2004)\citenamefont{Novoselov, Geim,
  Morozov, Jiang, Zhang, Dubonos, Grigorieva, and Firsov}}]{kn:novoselov2004}
\bibinfo{author}{\bibfnamefont{K.~S.} \bibnamefont{Novoselov}},
  \bibinfo{author}{\bibfnamefont{A.~K.} \bibnamefont{Geim}},
  \bibinfo{author}{\bibfnamefont{S.~V.} \bibnamefont{Morozov}},
  \bibinfo{author}{\bibfnamefont{D.}~\bibnamefont{Jiang}},
  \bibinfo{author}{\bibfnamefont{Y.}~\bibnamefont{Zhang}},
  \bibinfo{author}{\bibfnamefont{S.~V.} \bibnamefont{Dubonos}},
  \bibinfo{author}{\bibfnamefont{I.~V.} \bibnamefont{Grigorieva}},
  \bibnamefont{and} \bibinfo{author}{\bibfnamefont{A.~A.}
  \bibnamefont{Firsov}}, \bibinfo{journal}{Science}
  \textbf{\bibinfo{volume}{306}}, \bibinfo{pages}{666} (\bibinfo{year}{2004}).

\bibitem[{\citenamefont{Novoselov et~al.}(2005)\citenamefont{Novoselov, Geim,
  Morozov, Jiang, Zhang, Katsnelson, Grigorieva, Dubonos, and
  Firsov}}]{kn:novoselov2005}
\bibinfo{author}{\bibfnamefont{K.~S.} \bibnamefont{Novoselov}},
  \bibinfo{author}{\bibfnamefont{A.~K.} \bibnamefont{Geim}},
  \bibinfo{author}{\bibfnamefont{S.~V.} \bibnamefont{Morozov}},
  \bibinfo{author}{\bibfnamefont{D.}~\bibnamefont{Jiang}},
  \bibinfo{author}{\bibfnamefont{Y.}~\bibnamefont{Zhang}},
  \bibinfo{author}{\bibfnamefont{M.~I.} \bibnamefont{Katsnelson}},
  \bibinfo{author}{\bibfnamefont{I.~V.} \bibnamefont{Grigorieva}},
  \bibinfo{author}{\bibfnamefont{S.~V.} \bibnamefont{Dubonos}},
  \bibnamefont{and} \bibinfo{author}{\bibfnamefont{A.~A.}
  \bibnamefont{Firsov}}, \bibinfo{journal}{Nature}
  \textbf{\bibinfo{volume}{438}}, \bibinfo{pages}{197} (\bibinfo{year}{2005}).

\bibitem[{\citenamefont{Zhang et~al.}(2005)\citenamefont{Zhang, Tan, Stormer,
  and Kim}}]{kn:zhang2005}
\bibinfo{author}{\bibfnamefont{Y.}~\bibnamefont{Zhang}},
  \bibinfo{author}{\bibfnamefont{Y.-W.} \bibnamefont{Tan}},
  \bibinfo{author}{\bibfnamefont{H.~L.} \bibnamefont{Stormer}},
  \bibnamefont{and} \bibinfo{author}{\bibfnamefont{P.}~\bibnamefont{Kim}},
  \bibinfo{journal}{Nature} \textbf{\bibinfo{volume}{438}},
  \bibinfo{pages}{201} (\bibinfo{year}{2005}).


\bibitem{kn:hwang2006c}
E. H. Hwang, S. Adam, and S. Das Sarma, Phys. Rev. Lett {\bf 98},
186806 (2007); and arXiv:cond-mat/0610834 (2006).

\bibitem[{\citenamefont{Nomura and MacDonald}(2006)}]{kn:nomura2006a}
\bibinfo{author}{\bibfnamefont{K.}~\bibnamefont{Nomura}} \bibnamefont{and}
  \bibinfo{author}{\bibfnamefont{A.~H.} \bibnamefont{MacDonald}},
  \bibinfo{journal}{Phys. Rev. Lett.} \textbf{\bibinfo{volume}{96}},
  \bibinfo{pages}{256602} (\bibinfo{year}{2006});
\bibinfo{author}{\bibfnamefont{V.~V.} \bibnamefont{Cheianov}} \bibnamefont{and}
  \bibinfo{author}{\bibfnamefont{V.~I.} \bibnamefont{Fal'ko}},
  \bibinfo{journal}{Phys. Rev. Lett.} \textbf{\bibinfo{volume}{97}},
  \bibinfo{pages}{226801} (\bibinfo{year}{2006}).

\bibitem[{\citenamefont{Fradkin}(1986)}]{kn:fradkin1986}
\bibinfo{author}{\bibfnamefont{E.}~\bibnamefont{Fradkin}},
  \bibinfo{journal}{Phys. Rev. B} \textbf{\bibinfo{volume}{33}},
  \bibinfo{pages}{3257} (\bibinfo{year}{1986});
\bibinfo{author}{\bibfnamefont{A.~W.~W.} \bibnamefont{Ludwig}},
  \bibinfo{author}{\bibfnamefont{M.~P.~A.} \bibnamefont{Fisher}},
  \bibinfo{author}{\bibfnamefont{R.}~\bibnamefont{Shankar}}, \bibnamefont{and}
  \bibinfo{author}{\bibfnamefont{G.}~\bibnamefont{Grinstein}},
  \bibinfo{journal}{Phys. Rev. B} \textbf{\bibinfo{volume}{50}},
  \bibinfo{pages}{7526} (\bibinfo{year}{1994}).

\bibitem[{\citenamefont{Katsnelson}(2006)}]{kn:katsnelson2006}
\bibinfo{author}{\bibfnamefont{M.~I.} \bibnamefont{Katsnelson}},
  \bibinfo{journal}{Eur. Phys. J. B} \textbf{\bibinfo{volume}{51}},
  \bibinfo{pages}{157} (\bibinfo{year}{2006});
\bibinfo{author}{\bibfnamefont{J.}~\bibnamefont{Tworzyd\l{}o}},
  \bibinfo{author}{\bibfnamefont{B.}~\bibnamefont{Trauzettel}},
  \bibinfo{author}{\bibfnamefont{M.}~\bibnamefont{Titov}},
  \bibinfo{author}{\bibfnamefont{A.}~\bibnamefont{Rycerz}}, \bibnamefont{and}
  \bibinfo{author}{\bibfnamefont{C.~W.~J.} \bibnamefont{Beenakker}},
  \bibinfo{journal}{Phys. Rev. Lett.} \textbf{\bibinfo{volume}{96}},
  \bibinfo{pages}{246802} (\bibinfo{year}{2006}).


\bibitem{Nixon}
J. A. Nixon and J. H. Davis, Phys. Rev. B {\bf 41}, 7929 (1990).

\bibitem{GG}
B. Huard, J.A. Sulpizio, N. Stander, K. Todd, B. Yang, and D.
Goldhaber-Gordon, Phys. Rev. Lett. {\bf 98}, 236803 (2007).


\bibitem[{\citenamefont{Adam et~al.}(2007)\citenamefont{Adam, Hwang, Galitski,
  and {\mbox Das Sarma}}}]{AHGD}
\bibinfo{author}{\bibfnamefont{S.}~\bibnamefont{Adam}},
  \bibinfo{author}{\bibfnamefont{E.~H.} \bibnamefont{Hwang}},
  \bibinfo{author}{\bibfnamefont{V.~M.} \bibnamefont{Galitski}},
  \bibnamefont{and} \bibinfo{author}{\bibfnamefont{S.}~\bibnamefont{{\mbox Das
  Sarma}}}, \bibinfo{journal}{Proc. Natl. Acad. Sci. USA, in press
  (arXiv:0705.1540 [cond-mat.mes-hall])}  (\bibinfo{year}{2007}).

\bibitem{CFAA}
V.V. Cheianov, V.I. Falko, B.L. Altshuler, and I.L. Aleiner,
Phys. Rev. Lett {\bf 99}, 176801 (2007); and
arXiv:0706.2968 (2007).

\bibitem[{\citenamefont{Tan et~al.}(2007)\citenamefont{Tan, Zhang, Bolotin,
  Zhao, Adam, Hwang, \mbox{Das Sarma}, Stormer, and Kim}}]{Columbia}
\bibinfo{author}{\bibfnamefont{Y.-W.} \bibnamefont{Tan}},
  \bibinfo{author}{\bibfnamefont{Y.}~\bibnamefont{Zhang}},
  \bibinfo{author}{\bibfnamefont{K.}~\bibnamefont{Bolotin}},
  \bibinfo{author}{\bibfnamefont{Y.}~\bibnamefont{Zhao}},
  \bibinfo{author}{\bibfnamefont{S.}~\bibnamefont{Adam}},
  \bibinfo{author}{\bibfnamefont{E.~H.} \bibnamefont{Hwang}},
  \bibinfo{author}{\bibfnamefont{S.}~\bibnamefont{\mbox{Das Sarma}}},
  \bibinfo{author}{\bibfnamefont{H.~L.} \bibnamefont{Stormer}},
  \bibnamefont{and} \bibinfo{author}{\bibfnamefont{P.}~\bibnamefont{Kim}},
  \bibinfo{journal}{Phys. Rev. Lett. in press (arXiv:0707.1807v1
  [cond-mat.mes-hall])}  (\bibinfo{year}{2007}).

\bibitem[{\citenamefont{Larkin et~al.}(1971)\citenamefont{Larkin, Mel'nikov,
  and Khmelnitskii}}]{kn:larkin1971}
\bibinfo{author}{\bibfnamefont{A.~I.} \bibnamefont{Larkin}},
  \bibinfo{author}{\bibfnamefont{V.}~\bibnamefont{Mel'nikov}},
  \bibnamefont{and}
  \bibinfo{author}{\bibfnamefont{D.}~\bibnamefont{Khmelnitskii}},
  \bibinfo{journal}{Zh. Eksp. Teor. Fiz.} \textbf{\bibinfo{volume}{60}},
  \bibinfo{pages}{846} (\bibinfo{year}{1971}), \bibinfo{note}{[Sov.\ Phys.\
  JETP {\bf 33}, 458 (1971)]};
\bibinfo{author}{\bibfnamefont{V.~M.} \bibnamefont{Galitski}} \bibnamefont{and}
  \bibinfo{author}{\bibfnamefont{A.~I.} \bibnamefont{Larkin}},
  \bibinfo{journal}{Phys. Rev. B} \textbf{\bibinfo{volume}{66}},
  \bibinfo{pages}{064526} (\bibinfo{year}{2002}).

\bibitem[{\citenamefont{Das~Sarma et~al.}(2005)\citenamefont{Das~Sarma, Lilly,
  Hwang, Pfeiffer, West, and Reno}}]{kn:dassarma2005b}
\bibinfo{author}{\bibfnamefont{S.}~\bibnamefont{Das~Sarma}},
  \bibinfo{author}{\bibfnamefont{M.~P.} \bibnamefont{Lilly}},
  \bibinfo{author}{\bibfnamefont{E.~H.} \bibnamefont{Hwang}},
  \bibinfo{author}{\bibfnamefont{L.~N.} \bibnamefont{Pfeiffer}},
  \bibinfo{author}{\bibfnamefont{K.~W.} \bibnamefont{West}}, \bibnamefont{and}
  \bibinfo{author}{\bibfnamefont{J.~L.} \bibnamefont{Reno}},
  \bibinfo{journal}{Phys. Rev. Lett.} \textbf{\bibinfo{volume}{94}},
  \bibinfo{pages}{136401} (\bibinfo{year}{2005}).

\bibitem{kn:martin2007}
J. Martin, N. Akerman, G. Ulbricht, T. Lohmann, J. H. Smet, K.
{von~Klitzing}, and A. Yacoby, arXiv:0705.2180 (2007).


\end{thebibliography}
\end{document}